\documentclass[11pt,a4paper]{article}
\usepackage{jheppub}

\title{\boldmath 
Probing  Sterile Neutrino via Lepton Flavor Violating Decays of Mesons}

\author{Shiyong Hu, Sam Ming-Yin Wong,}
\author{Fanrong Xu\footnote{Corresponding author.}}

\affiliation
{Department of Physics and Siyuan Laboratory, Jinan University,\\
$\;$Guangzhou 510632, P.R. China}

\emailAdd{fanrongxu@jnu.edu.cn}

\abstract{
A sterile neutrino at GeV mass scale is of particular interest in this work.
Though not take part in neutrino oscillation, the sterile neutrino 
can induce flavor violating semileptonic and leptonic decay of $K, D$ and $B$
mesons. We calculated a box diagram contribution in these processes.
By making use of current experiment limit of
lepton flavor violating decays $M^+_h\to M_l^+\ell_1^+\ell_2^-$ and $M^0\to \ell_1^-\ell_2^+$,
we explore the allowed parameter space of 
$U_{e4}$ and $U_{\mu 4}$ in different mass ranges.
Generally speaking,  both channels give a limit to the product of $U_{e4}$ and  $U_{\mu 4}$.
When sterile neutrino mass is located in between
pion and kaon mass, $K^+\to \pi^+ e^{\pm}\mu^{\mp}$ gives the strongest constraint 
while $B^+\to \pi^+ e^{\pm} \mu^{\mp}$
provides the dominated constraint when its mass in between of kaon and B meson. 
If sterile neutrino is even heavier than B mesons, 
the  $B^0_s\to \mu^{\pm} e^{\mp}$ experiment which is performed at LHCb
gives the strongest constraint. 
 }


\begin{document}
\maketitle
\flushbottom


\section{Introduction}
\label{sec:Intro}



Now it has been well established that at least two active neutrinos are massive with tiny masses. 
The origin of neutrino mass is still an open question. Sorts of ideas have been proposed to solve
this fundamental question, including seesaw mechanism \cite{seesaw} and radiative correction mechanism (for example,\cite{Kohda:2012sr}, \cite{Chang:2016zll} and for a recent review see \cite{Cai:2017jrq}).
General speaking, new particles out of SM particle spectrum will appear associated with neutrino mass models.
As a hypothetic particle, though does not participate weak interaction, sterile neutrino 
is unavoidable  in some neutrino mass models beyond SM.
For example, in Type I seesaw mechanism the heavy right-handed neutrino singlet contributing
the tiny mass of left-handed neutrino is  absent from SU(2) interaction
and hence  appears as a sterile neutrino.


The prediction of sterile neutrino mass in theory is model dependent thus is not unique.
In the view of experiment, there are some hints to indicate the existence of sterile neutrino
as well as its mass. One type of experiment is neutrino oscillation. 
In 2001 the LSND experiment searched $\bar{\nu}_\mu\to\bar{\nu}_e$ oscillations,
suggesting that neutrino oscillations occur in the $0.2<\Delta m^2<10\, {\rm{eV}}^2$  range \cite{Aguilar:2001ty}.
Later the MiniBooNE experiment indicated a two-neutrino oscillation, $\bar{\nu}_\mu\to\bar{\nu}_e$,
occurred in the $0.01<\Delta m^2<1.0\,{\rm{eV}}^2$ range \cite{Aguilar-Arevalo:2013pmq}.
An updated global fit \cite{Gariazzo:2017fdh}, taking into account recent progress, gives
$\Delta m_{41}^2\approx 1.7  {\rm{eV}}^2$ (best-fit), $1.3  {\rm{eV}}^2$ (at 2$\sigma$),
$2.4  {\rm{eV}}^2$ (at $3\sigma$).
Hence  if its  mass is  located at eV,  sterile neutrino effect can be unfolded by oscillation experiments. 
On the other hand, sterile neutrino mass can be even heavier.
The operation of LHC provides an opportunity 
to search TeV scale heavy sterile neutrino \cite{Pascoli:2018heg}, \cite{Atre:2009rg}.
The IceCube Neutrino Observatory, which locates in Antarctic, gives an unique vision 
to observe PeV neutrino \cite{IceCube:2018}.  
In between eV and TeV-PeV, the GeV sterile neutrino could appear 
in weak decays of bound states of heavy quarks.
Thus the sterile neutrino, with its undermined mass varied from eV to TeV-PeV,
provides a port to connect New Physics beyond SM.

In recent years the semileptonic decays $B^0\to K^{*}\ell^+\ell^-$ and $B^0\to \pi \ell^+ \ell^-$  have been studied 
extensively both theoretically and experimentally. Though the expectation of new physics in forward-backward
asymmetry of lepton pairs in $B^0\to K^{*}\ell^+ \ell^-$ has already faded away,  NP hope still holds at the
so-called observable $P'_5$ \cite{Aaboud:2018krd, Sirunyan:2017dhj}.  Similar situation happened in the leptonic decays of $B^0$ and $B_s^0$.
The SM-like $\mathcal{B}(B_s\to\mu^+\mu^-)$ and perhaps new physics allowed $\mathcal{B}(B_d\to\mu^+\mu^-)$
give strong constraints to theories beyond SM \cite{Hou:2013btm, Hou:2014nna}, but the windows to NP is not closed. 
It is known that both types of decays are FCNC process, giving a chance to put NP particles in the loop.
Then it is nature to consider the possibility of a sterile neutrino in the loop.
In fact there have been continuous efforts  to study GeV scale sterile neutrino indirectly via some certain semileptonic and leptonic  decays of  $B, D$ and $K$ mesons. 
In the semileptonic decay processes, if its mass is in between the meson masses of initial and final 
states,  the sterile neutrino can be on-shell produced \cite{Cvetic:2016fbv}. 
A popular consideration is to take sterile neutrino as the Majorana neutrino, thus lepton number violating decays are induced\cite{Cvetic:2015naa},\cite{Cvetic:2010rw},
\cite{Milanes:2018aku},\cite{Zuber:2000vy},\cite{Wang:2014lda},\cite{Helo:2010cw}. 
The idea to make use of 
leptonic decay with neutrino final state, in which
 sterile neutrino is involved at tree level,
is also proposed\cite{Abada:2013aba}, \cite{Yue:2018hci}. 
In above works the final state leptons, though with lepton number violation, are mostly with same flavors thus only
single PMNS matrix is relevant. 
The lepton flavor violating decays from mesons, on the other hand, is related 
to two PMNS matrix elements, thus  could
give complementary information to corresponding lepton number violating decays.

The early quest for lepton flavor violating processes can be traced back to the
leptonic cecay $K_L\to e^{\pm}\mu^{\mp}$ in 1998
 \cite{Ambrose:1998us}. So far $K_L \to e^{\pm} \mu^{\mp}$  still gives a very strong constraint
 to NP models. 
The latest experiment for leptonic decay is carried out in LHCb by searching
$B^0\to e^{\pm} \mu^{\mp}$ giving the upper limit $1.3\times 10^{-9}$ \cite{Aaij:2017cza}.
As for the  semileptonic decays of $K, D$ and $B$ mesons, most of them are still
results from BaBar \cite{Aubert:2007mm} and it is hoped that LHCb can bring new limit in the near future.
A detailed summary for related experiments is given in Table \ref{tab:status}.
In this paper, we will analysis both leptonic and semileptonic decays from $K, D$ and $B$ mesons induced
by sterile neutrino. 
By combing all the currently related experiments, we will give the constraints to relevant PMNS matrix 
elements.

This paper is organized as follows. In section \ref{sec:model},
 we will give a brief introduction to the model related to heavy sterile neutrino.
In section \ref{sec:semi} we will discuss a set of semileptonic decay processes 
and derive the exact formulas with heavy sterile neutrino contribution.
A systematic formalism for leptonic decays of $K, D$ and $B$ mesons are 
given in section \ref{sec:lep}.
In section \ref{sec:num}, we will perform a numerical study 
and give the allowed 
coherent parameter space. 
Discussion and conclusion will be made in section \ref{sec:con}. 

\section{Working Frame}

In this section, a brief introduction of heavy sterile neutrino is given firstly.
Then we  derive the required analytical formulas in semileptonic
and leptonic decays separately. 

\subsection{Model setting}\label{sec:model}

As introduced in section \ref{sec:Intro}, here we are only interested in the 
GeV scale sterile neutrino. 
Due to its heavy mass, in flavor space the sterile neutrino will decouple from 
other three active neutrinos in the oscillation processes. With
the appearance of a sterile neutrino and without involving the details of
a concrete model, the mass mixing can always
written via a non-unitary mixing matrix,
\begin{equation}
\left(
\begin{array}{c}
\nu_e\\ \nu_\mu\\ \nu_\tau \\
\end{array}\right)
=\left(\begin{array}{cccc}
U_{e1} & U_{e2} & U_{e3} & U_{e4}\\
U_{\mu1} & U_{\mu 2} & U_{\mu 3} & U_{ \mu 4}\\
U_{\tau 1} & U_{\tau 2} & U_{\tau 3} & U_{\tau 4}
\end{array}
\right)
\left(
\begin{array}{c}
\nu_1\\ \nu_2\\ \nu_3 \\ \nu_4
\end{array}\right),\label{eq:mixing}
\end{equation}
which characterizes the rotation between mass eigenstate and flavor eigenstate in vacuum.
A direct consequent for the non-unitary mixing is zero-distance effect \cite{Li:2015oal}, 
the oscillation could happen even without propagate few distance. 
Such effect has been pointed out to be detected in oscillation experiment by a near detector,
which will be discussed in a separate work. 
In the following context, we will focus on the mass of $\nu_4$ and the mixing elements $U_{e4,\mu 4, \tau 4}$.
And hereafter we adopt the notation $N$ to denote sterile neutrino for the purpose of emphasize.


\subsection{Semileptonic decay}
\label{sec:semi}

\begin{figure}[h]
\begin{center}
\includegraphics[width=8cm]{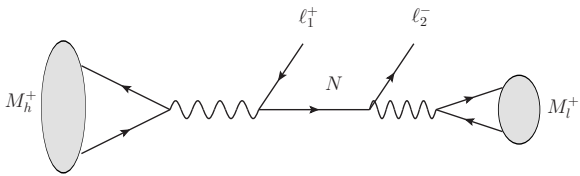}
\caption{The semileptonic decay of charged heavy mesons, in which $M_h^+ (M_l^+)$
means a  charged heavy (light) meson.}
\label{fig:semileptonic}
\end{center}
\end{figure}

The sterile neutrino, if its mass is in between with the initial heavy  meson and final mass meson,  can be produced on-shell and then decay shortly. 
As for the heavy meson,
we are especially interested in those charged ones.
The  reason for such a choice is due to the fact that
 tree-level annihilation diagram not only gives dominated contribution to the decay of heavy meson,
 but also provides a chance to produce sterile neutrino from $W$ boson  sourced from
 quark annihilation, see Fig. \ref{fig:semileptonic}. 

The branching fraction for the three-body decay  can be simplified to 
the multiplication of  two-body decays.
In general we can write down
\begin{equation}
\mathcal{B}(M_h^+\to M_l^+ \ell_1^+ \ell_2^-)=\mathcal{B}(M_h^+\to \ell_1^+ N) \mathcal{B}
(N\to M_l^+ \ell_2^-)\label{eq:3-body}
\end{equation}
in which $M_{h} (M_{l})$ denotes heavy (light) meson, 
 and $\ell_{1}, \ell_2$ represent charged lepton ($e, \mu, \tau$) with different flavors.
A straightforward calculation gives the decay of heavy meson 
\begin{equation}
\mathcal{B}(M_h^+ \to \ell_1^+ N)=\frac{1}{8\pi}G_F^2 f_{M_h}^2 m_{M_h} m_N^2 \tau_{M_h}|U_{\ell_1 4}|^2 
X(M_h).
\label{eq:2-body-1}
\end{equation}
which relies on  two unknown parameters, the PMNS matrix element $U_{\ell_1 4}$ and the mass of $N$.  
Especially the heavy meson dependent function $X(M_h)$ introduced in Eq. (\ref{eq:2-body-1}) reflects
the features of $M_h$,
\begin{equation}
X(M_h)=|\xi_1|^2\lambda^{\frac12}(1,x_N, x_{\ell_1})[(1+x_{\ell_1}-x_N)  + y_{\ell_1}(1+x_N-x_{\ell_1}) ].
\label{eq:X}
\end{equation}
in which $\xi_1$ is a particular  CKM matrix element 
corresponding to the mother particle,  and the definition of the  auxiliary function  is given as
$\lambda(x,y,z)=x^2+y^2+z^2-2(xy+xz+yz)$.
The two parameters correlated to initiated and final state particle are defined as
 $x_i\equiv \frac{m_i^2}{m_{M_h}^2}, y_i\equiv \frac{m_i^2}{m_N^2}$.

For the further decay of  sterile neutrino, one can calculate its decay width, 
\begin{equation}
\Gamma(N\to M_l {\ell_2^-})=\frac{1}{16\pi} G_F^2  m_N^3 |U_{\ell_2 4}|^2 Y(M_l).
\label{eq:2-body-2}
\end{equation}
Similar to the decay of $M_h$, 
besides the PMNS matrix $U_{\ell_2 4}$,  the width depends on the final state dependent function $Y$, given
\begin{equation}
Y(M_l, {\ell_2})=|\xi_2|^2 f^2_{M_l} \lambda^{\frac12}(1,y_{\ell_2}, y_{M_l}) [(1+y_{\ell_2}-y_{M_l})(1+y_{\ell_2})-4y_{\ell_2}],
\end{equation}
with another CKM matrix element $\xi_2$ which is determined by $M_l$.
The lepton final state from $W$ is assumed negligible, accordingly the branching fraction of $N$ decay is 
\begin{equation}
\mathcal{B}(N\to M_l \ell_2^-)=\frac{|U_{\ell_2 N}|^2 Y(M_l,m_{\ell_2})}
{\sum\limits_{\ell; q} |V_{\ell N}|^2 |V_{uq}|^2 Y(M_q, m_\ell)}
\end{equation}
In the denominator, the summation is performed only to the first two generations for both lepton and 
quark sector. As for the function $Y$, the value for its first parameter $M_q$ should
be chosen as $M_q= \pi^+ (K^+)$.

\begin{table}[h]
 \caption{The detailed parameters for semileptonic decay $M_h^+\to M_l^+\ell_1^+\ell_2^-$, in which 
 $M_h(M_l)$ means heavy (light) meson and $\ell_{1,2}=e,\mu, \tau$.
 } \label{tab:mesons}
 \vspace{0.3cm}
\begin{center}\label{tab:lep}
\begin{tabular}{ ccccc c|ccc}
\hline
$M_h^+$  & $\xi_1$  & $f_{M_h}$ & $m_{M_h}$  & $\tau_{M_h}$ & $x_i$& $M_l^+$      & $\xi_2$ &$f_{M_l}$\\
\hline\hline
$B^+$ &$V_{ub}$   & $f_{B}$ & $m_{B^+}$ & $\tau_{B^+}$  &$\frac{m_i^2}{m_{B^+}^2}$ & $K^+$   & $V_{us}$ &$f_K$\\
&  &&& &&$\pi^+$     &$V_{ud}$       &$f_\pi$         \\
\hline
$D^+$ &$V_{cd}$   & $f_D$ & $m_{D^+}$ & $\tau_{D^+}$  &$\frac{m_i^2}{m_{D^+}^2}$ & $K^+$      &$V_{us}$& $f_K$\\
&    & & &  &  & $\pi^+$     & $V_{ud}$     & $f_\pi$   \\
\hline
$K^+$ &  $V_{us}$  &$f_K$  &$m_{K^+}$  & $\tau_{K^+}$  &  $\frac{m_i^2}{m_{K^+}^2}$ &$\pi^+$  &$V_{ud}$  & $f_\pi$    \\
 \hline
\end{tabular}
\end{center}
\end{table}

One should keep in mind that Eq. (\ref{eq:3-body}) gives a general description of this type process, 
which actually contains many modes when different initial and 
final states are chosen. In Table \ref{tab:mesons} we have summarized explicitly corresponding parameters
for such modes.


\subsection{Leptonic decay}
\label{sec:lep}

The leptonic decay with different final state flavors in SM
is induced by active neutrino in box diagram, however, 
its effect is very tiny thus  can be negligible. 
On the other hand the smallness in SM might be made use of searching  new physics.
If the processes can be observed in experiment, it will be definitely a signal for desired new physics.
As an illustration, we will consider such a process induced by heavy sterile neutrino.

\begin{figure}[h]
\begin{center}
\includegraphics[width=8cm]{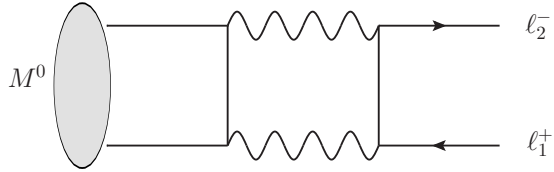}
\caption{Flavor violating leptonic decay of neutral mesons.}
\label{fig:lepton}
\end{center}
\end{figure}

To calculate the amplitude of usual final states with same flavor int theory, we will include 
 both penguin diagram and box diagram contribution.
In this work we only focus on the final state leptons with different flavor,  thus only box diagram contributes
since the vector boson in penguin diagram does not change flavor, see Fig. \ref{fig:lepton}. 
The initial neutral meson $M^0$, could be $B^0, D^0$ and  $K_L$ and the intermediate particle $N$ is off-shell.  
Generically we have the branching fraction for such decays
\begin{equation}
\mathcal{B}(M^0\to \ell_1^+ \ell_2^-)=
\frac{G_F^2\alpha^2 }{32\pi^3 \sin^4\theta_W}
A^2(z_h,z_N)  |U_{\ell_1 N}|^2 |U_{\ell_2N}|^2 Z(M^0, \ell_1,\ell_2)
\label{eq:lepton}
\end{equation}
which relies on PMNS matrix $U_{\ell_1 N}, U_{\ell_2 N}$ and a function correlated
with initial and final states
\begin{equation}
Z(M^0, \ell_1,\ell_2) \equiv |\xi|^2\tau(M^0) m_M^3
 f_M^2 \lambda^{\frac12}(1, x_{\ell_2}, x_{\ell_1})
\left[ x_{\ell_2}(1+x_{\ell_1} -x_{\ell_2})+x_{\ell_1}(1+x_{\ell_2}-x_{\ell_1})\right]
\end{equation}
with the product of relevant CKM elements $\xi$. The other relevant parameter $x_i$
is defined as $z_i\equiv \frac{m_i^2}{m_{W}^2}$. All the explicit parameters involved in
different decay modes are summarized in Table \ref{tab:lep}. 

\begin{table}[h]
 \caption{The detailed parameters for semileptonic decay $M^0\to \ell_1^+ \ell_2^-$. 
 } \label{tab:lep}
 \vspace{0.3cm}
\begin{center}\label{tab:lep}
\begin{tabular}{ ccc ccc}
\hline
$M^0$ &  $\xi(M^0)$     & $\tau(M^0)$ & $f_M$ & $m_M$  & $z_h$ \\
\hline\hline
$B^0_s$ &  $V_{tb}^*V_{ts}$   & $\tau(B_s^0)$ & $f_{B_s}$ & $m_{B_s^0}$ & $z_t$ \\
$B^0$ &  $V_{tb}^*V_{td}$   & $\tau(B^0)$ & $f_B$ & $m_{B^0}$ & $z_t$ \\
$D^0$ &  $V_{cb}^*V_{ub}$    & $\tau(D^0)$ & $f_D$ & $m_{D^0}$ & $z_b$\\
$K_L$ &  $V_{ts}^*V_{td}$    & $\tau(K_L)$ & $f_K$ & $m_{K_L}$ & $z_t$\\
 \hline
\end{tabular}
\end{center}
\end{table}

The loop function  $A(x,y)$ in Eq.(\ref{eq:lepton}), which is obtained by calculating  the box diagram in Fig. \ref{fig:lepton},
\begin{equation}
A(x, y)=\frac14\left[\frac{x -y}{(1-x)(1-y)}
+\frac{(1-y)^2 x^2 \ln x -(1-x)^2 y^2 \ln y}
{(1-x)^2(1-y)^2 (x-y)}\right], \label{eq:A}
\end{equation}
reveals the inner information of $M^0$.
It can be checked that function $A(x,y)$ is an extension of standard loop function $
B_0(x)=\frac14 \left[ \frac{x}{1-x}+\frac{x \ln x}{(x-1)^2}\right]$, and can return 
to $B_0$ case when the second parameter vanishes. 
A more qualitative
analysis for $A(x,y)$ will be given in next section.


\section{Numerical results and discussion}
\label{sec:num}

\subsection{Experiment status}

There have been about 20 years history for the  search for flavor violating decays.
We summarize all the relevant experiments in Table \ref{tab:status} as the input of 
our numerical study.

\begin{table}[h]
\centering  
\caption{The status of flavor violating semileptonic or leptonic decays related
to $K, D$ and $B$. The abbreviations are as follows: LED means light eigenstate dominate
and HED is heavy eigenstate dominate.}  
\label{tab:status}
\vspace{0.3cm}
\begin{tabular}{c|c|c}  
\hline
channel &  $90\%$ CL limits & collaboration  \\\hline\hline
$K^+\to \pi^+\mu^+ e^-$ & $1.3\times 10^{-11}$   &  \cite{Sher:2005sp} \\\
$K^+\to \pi^+\mu^- e^+$ & $5.2\times 10^{-10}$    &  \cite{Appel:2000tc}  \\\hline
$K_L^0\to  e^{\pm} \mu^{\mp}$ &$0.47\times 10^{-11}$  & BNL  \cite{Ambrose:1998us} \\ \hline
$D^+\to \pi^+\mu^+ e^-$ & $3.6\times 10^{-6}$ &  BaBar \cite{Lees:2011hb} \\ 
$D^+\to \pi^+\mu^- e^+$ & $2.9\times 10^{-6}$ & BaBar \cite{Lees:2011hb} \\ 
$D^+\to K^+\mu^+ e^-$ & $2.8\times 10^{-6}$ &  BaBar \cite{Lees:2011hb} \\
$D^+\to K^+\mu^- e^+$ &$1.2\times 10^{-6}$  &  BaBar \cite{Lees:2011hb}\\
\hline
$D^0\to e \mu$  & $1.3\times 10^{-8}$    &  LHCb \cite{Aaij:2015qmj}\\
\hline  
$B^+\to \pi^+ e^{\pm}\mu^{\mp}$ & $1.7\times 10^{-7}$ & BaBar \cite{Aubert:2007mm} \\
$B^+\to \pi^+ e^{\pm}\tau^{\mp}$ & $7.5 \times 10^{-5}$ & BaBar \cite{Lees:2012zz} \\
$B^+\to \pi^+\mu^{\pm} \tau^{\mp}$ & $7.2 \times 10^{-7}$ & BaBar \cite{Lees:2012zz} \\\hline
$B^+\to K^+e^{\pm} \mu^{\mp}$ & $9.1\times 10^{-8}$ & BaBar \cite{Aubert:2006vb} \\
$B^+\to K^+e^{\pm} \tau^{\mp}$ & $3.0\times 10^{-5}$ & BaBar \cite{Lees:2012zz}\\
$B^+\to K^+\mu^{\pm} \tau^{\mp}$ &$4.8\times 10^{-5}$   & BaBar \cite{Lees:2012zz} \\ \hline
$B^+\to K^{*+}e^{\pm} \mu^{\mp}$ &$1.4\times 10^{-6}$  & BaBar \cite{Aubert:2006vb} \\ \hline
$B^0\to  e^{\pm} \mu^{\mp}$ &$2.8\times 10^{-9}$  &  LHCb \cite{Aaij:2013cby} \\ 
$B^0\to  e^{\pm} \tau^{\mp}$ &$2.8\times 10^{-5}$  & BaBar \cite{Aubert:2008cu}  \\ 
$B^0\to  \mu^{\pm} \tau^{\mp}$ &$2.2\times 10^{-5}$  & BaBar \cite{Aubert:2008cu} \\ \hline
$B^0\to e^{\pm} \mu^{\mp}$  & $1.3\times 10^{-9}$    &   LHCb \cite{Aaij:2017cza}\\
$B^0_s\to e^{\pm} \mu^{\mp}$  & $6.3\times 10^{-9}$   (LED) &   LHCb \cite{Aaij:2017cza}\\
$B^0_s\to e^{\pm} \mu^{\mp}$  & $7.2\times 10^{-9}$   (HED) &   LHCb \cite{Aaij:2017cza}\\\hline
\end{tabular}  
\end{table} 


\subsection{Property of $A(x,y)$}

The branching fractions of leptonic decays largely relies on the $A(x,y)$, hence before 
numerical studies of phenomenology  it is necessary 
to  explore the features of this function. In Fig.\ref{fig:lepton} we plot the dependence 
its behaviors respect to sterile neutrino mass.  
Typic features of $A(x,y)$ are shown below.
\begin{itemize}
\item A singularity appears at $m_N=m_W$, and more  close to $W$ mass,  more enhanced the function
value is.
\item In SM such a diagram actually also gives contribution, given $A\sim -0.2$ when neutrino 
mass is tiny.
\item There is a particular choice that $A=0$. Take B decay as an example, the internal heavy quark loop
comes from top. And the zero point is located at top quark mass region.  However, in SM such an effect is 
not appear as this Feynman diagram does not appear individually. 

\item When sterile neutrino mass is larger than electroweak scale, the behavior is asymptotic stable, 
giving a value smaller than SM. Since we are only interested in GeV scale sterile neutrino,
such a range will not be involved in this paper.

\end{itemize}

\begin{figure}[h]
\begin{center}
\includegraphics[width=10cm]{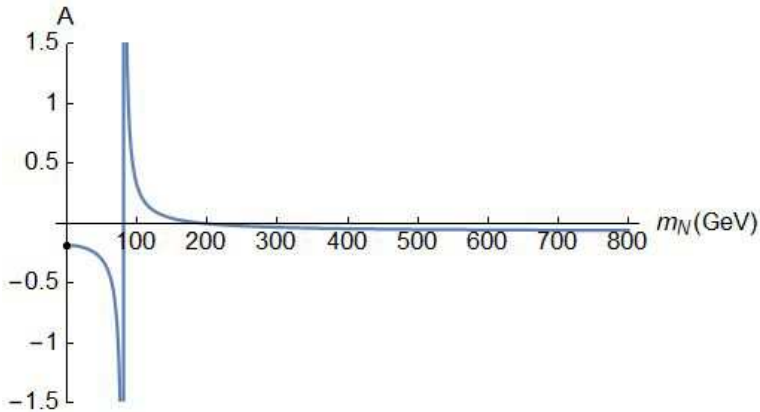}
\caption{The behavior of function $A$, in which the black dot stands for SM situation.}
\label{fig:lepton}
\end{center}
\end{figure}


\subsection{The combining analysis}


Now with the prepared necessary analytical formulas in above, we will present our 
numerical study in the following section. 

The semileptonic decay happens, in our working frame, is due to the on-shell production of
sterile neutrino, which actually requires the sterile neutrino mass in between of the initial heavy meson
and the final light meson. However, the effect of off-shell sterile neutrino can play a role in leptonic decays.  
Thus whatever mass of sterile neutrino is, the contribution from leptonic decays cannot be negligeble.
In other words, if the mass is not located in between initial and final mesons, only the leptonic decay experiments
give constraints to corresponding mixing matrix, we call this scenario D. In addition to scenario D, in the mass region
$m_\pi <m_N<m_B$, we classify the mass range into three other different scenarios, named as
Scenario A, B and C.

\begin{itemize}
\item Scenario A: $m_\pi < m_N<m_K$

If sterile neutrino mass located at this region, the semileptonic decays induced by the on-shell sterile neutrino contains 
$B^+\to \pi^+ \mu^{\pm} e^{\mp}, B^+\to \pi^+ \tau^{\pm} \mu^{\mp}, B^+\to \pi^+ \tau^{\pm} e^{\mp}, D^+\to \pi^+ \mu^{\pm} e^{\mp}$ and $ K^+\to \pi^+ \mu^{\pm} e^{\mp}$. In principle, all the leptonic decays from $K_L, D^0 $ and $B^0, B^0_s$,
including $K_L\to \mu^{\pm} e^{\mp}, D^0\to \mu^{\pm} e^{\mp}$ and $B_{(s)}^0\to \mu^{\pm} e^{\mp}$, 
should also be taken into account.  However, from the numerical analysis all the parameter spaces are fully allowed by these leptonic decays, which is too weak to give an efficient constraint.  Thus only these semileptonic ones 
provide some effective information.

Taking $\mu, e$ final states as an example, we compare the decays from three different parent particle and find $K^+$ decay 
provides the most stringent constraint shown in Table \ref{fig:scA}, while $B^+$ and $D^+$ decays give a much wide allowed region.  It is easily to see the product of $U_{eN}$ and $U_{\mu N}$ is strictly constrained to $\mathcal{O}(10^{-5})$, but 
a further restriction to $U_{\alpha N}$ requires other input experiment, which will be discussed in a separate work.
 
 \begin{figure}[h]
\begin{center}
\includegraphics[width=12cm]{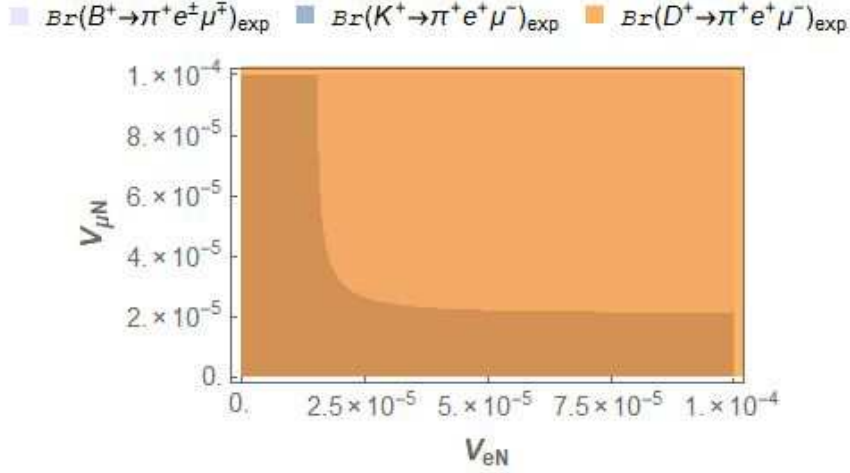}
\caption{The allowed parameter space in scenario A, in which we have taken $m_N=0.3 \,{\rm {GeV}}$ as 
an illustration.}
\label{fig:scA}
\end{center}
\end{figure}

Similar analysis can be performed for $\mu, \tau$ or $\tau, e$ final states. The output of the correlated constraint for relevant 
PMNS matrix elements is somehow too weak, thus we will not show them explicitly.

 \item Scenario B: $m_K < m_N<m_D$
 
 As pointed in above context, leptonic decays always appear. For the semileptonic decays in this case, only
 $B^+$ and $D^+$ decays while $K^+$ is forbidden otherwise the mother particle is lighter than its
  daughter particle. The explicit modes which are incorporated into our numerical simulations are:
 $B^+\to K^+(\pi^+) \mu^{\pm} e^{\mp}, B^+\to K^+(\pi^+) \tau^{\pm} \mu^{\mp}, B^+\to K^+(\pi^+) \tau^{\pm} e^{\mp}$
 and $D^+\to K^+(\pi^+) \mu^{\pm} e^{\mp}$.
 
As the first step, let's focus on $e,\mu$ final states.
First by comparing various $B^+$ decay modes with different final states, one can find the constraint to
PMNS matrix from $B^+\to \pi^+ e^{\pm} \mu^{\mp}$ dominates the corresponding ones from 
$B^+\to K^+ e^{\pm} \mu^{\mp}$, as shown in Table \ref{fig:scB}. Second
for the allowed region extracted from $D^+$ decays, $D^+\to \pi^+ e^+ \mu^-$ is much stronger than 
$D^+\to K^+ e^{\pm}\mu^{\mp}$.  Looking at the same $\pi^+$ final states, the numerical analysis tells
that $B^+$ decay gives the strongest restriction,
which actually gives an upper limit for the product of $U_{eN}$ and $U_{\mu N}$, $\mathcal{O}(10^{-2})$.
As for the individual matrix elements, one has to resort to other way.
 
 \begin{figure}[h]
\begin{center}
\includegraphics[width=7.8cm]{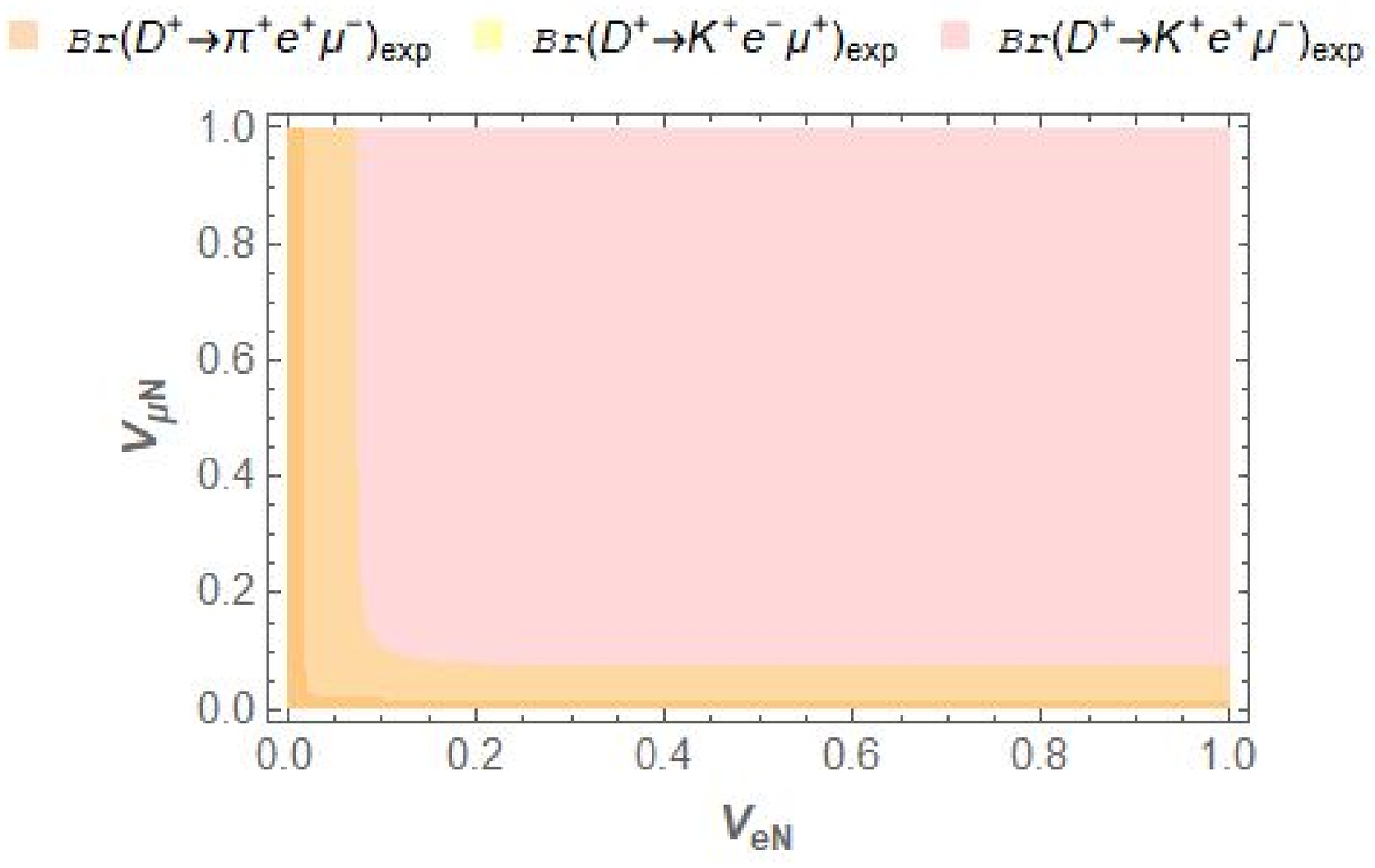}
\includegraphics[width=7cm]{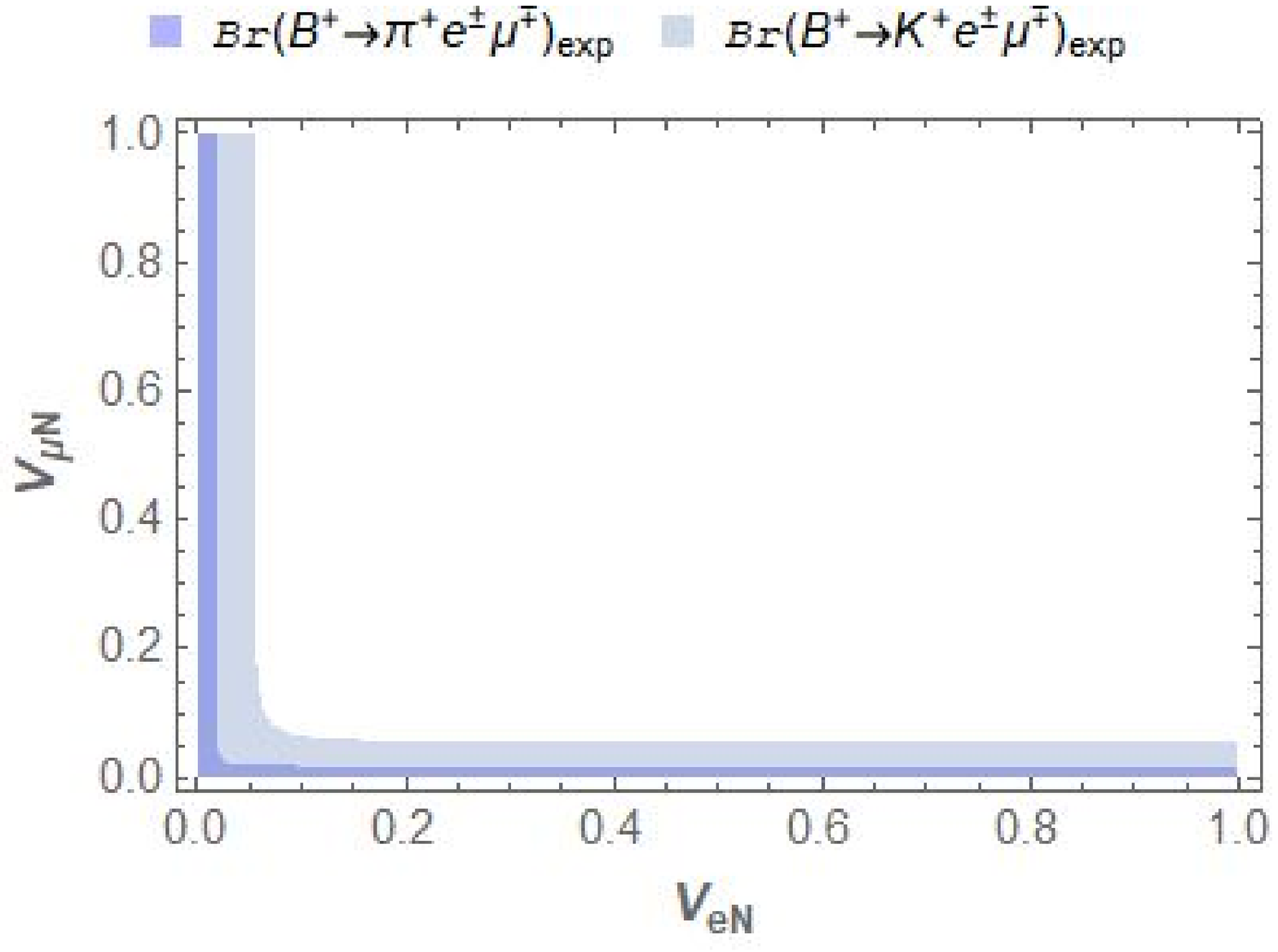}
\caption{The allowed parameter space in scenario B, in which we have taken $m_N=1.5 \,{\rm {GeV}}$ as 
an illustration.}
\label{fig:scB}
\end{center}
\end{figure}

It is noted that so far no more stringent constraint can be obtained from leptonic decay.
And the constraint of the correlation of PMNS matrix $V_{\tau N}$ and $V_{eN}, V_{\mu N}$ is still too weak 
from $\tau, \mu$ or $\tau, e$ final states, which is also neglected here.

 \item Scenario C: $m_D < m_N<m_B$
 
 In addition to leptonic decays, only semileptonic decays from $B^+$ can appear in this situation, which actually
 gives more stringent constraints.
 
 We still stick to the $\mu, e$ final states with the same reason as previous scenarios.
 Though sterile neutrino mass is taken $4\,{\rm{GeV}}$, the numerical simulation leads
 to the same conclusion as scenario B. Thus  we will not show  its corresponding plot here.



 \item Scenario D: $m_N<m_\pi$ or $m_N>m_B$
 
 In this scenario, semileptonic decays are forbidden and only leptonic decays happen.
 If the sterile neutrino mass is lighter than the lightest meson $\pi$, the mass dependent function $A$ is 
 close to the SM situation giving a small amplitude (module to PMNS matrix element), then
 the further constraint to PMNS matrix from experiment measurement is weak.
 Such behavior has been checked and we will not show in graphs here.

 \begin{figure}[h]
\begin{center}
\includegraphics[width=10cm]{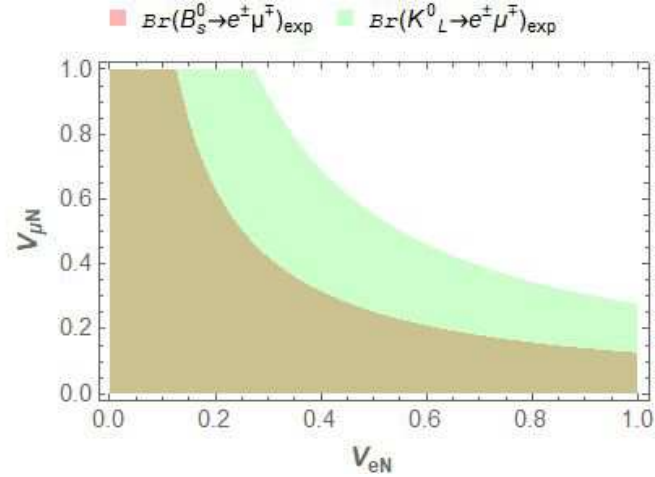}
\caption{The allowed parameter space in scenario D, in which we have taken $m_N=70 \,{\rm {GeV}}$ as 
an illustration.}
\label{fig:scD}
\end{center}
\end{figure}

It is interesting to explore the mass range larger than $B$ mesons. We take Fig. \ref{fig:scD}
 as an illustration, with sterile neutrino mass $m_N=70\,{\rm GeV}$.  Among all the 4 charged neutral
 meson decays, the parameter space from $D^0$ and $B^0$ decays are still fully filled thus 
 not marked in the figure. The experiment upper limit for $K_L$ and $B_s^0$ 
 indeed touch the  restriction to $U_{eN}-U_{\mu N}$ parameter space. 
 As shown in Fig. \ref{fig:scD}, the recent LHCb experiment $B_s^0\to e^{\pm} \mu^{\mp}$ now
 catch up with the classical BNL experiment on $K_L\to e^{\pm}\mu^{\mp}$.

\end{itemize}

In a summary, the allowed parameter of PMNS matrix is sterile neutrino mass dependent. 
When the mass is lighter than $m_\pi$, parameter space does not receive a 
constraint from current meson decay experiments. If the mass is located between
$m_\pi$ and $m_K$, the semileptonic decay $K^+\to \pi^+ e^+ \mu^-$ provides the
most stringent constraint, $V_{\mu N} V_{eN}\sim \mathcal{O}(10^{-5})$. When its mass
is in between kaon mass and $B$ meson, BaBar experiment $B^+\to \pi^+e^{\pm} \mu^{\mp}$ in fact 
dominate the constraint, giving $V_{\mu N} V_{eN}\sim \mathcal{O}(10^{-2})$.
If sterile neutrino is heavier than B meson, leptonic decay $B_s^0\to e^{\pm} \mu^{\mp}$ provides the strongest
constraint.


\section{Conclusion}
\label{sec:con}
\label{sec:conclusion}
In this work, we consider various semileptonic and leptonic decays of neutral mesons induced by a heavy
sterile neutrino, which can in turn constrain parameter space
of the unknown PMNS matrix elements. 
Especially we calculated the loop function of a box diagram contributing to leptonic decays.
Making use of the two types of decays from different parent particle, we find the allowed range of 
parameter space of  sterile neutrino is mass dependent. If sterile neutrino is lighter than pion mass,
these meson decays have null restriction. When sterile neutrino mass is located in between
pion and kaon mass, $K^+\to \pi^+ e^{\pm}\mu^{\mp}$ gives the strongest constraint while $B^+\to \pi^+ e^{\pm} \mu^{\mp}$
provides the dominated constraint when sterile neutrino mass in between of kaon and B meson. 
If sterile neutrino is even heavier than B mesons, the measurement performed at LHCb $B^0_s\to \mu^{\pm} e^{\mp}$
gives the strongest constraint.  It should be noted that so far we can only extract
the restriction information to parameter space from  the decays with $e,\mu$ final states while the decays
with a $\tau$ in final state is not incorporated. From the analysis, we provide a global constraint for
$|V_{\mu N} V_{eN}|$ in different $m_N$ mass region, however, the magnitude of an individual 
PMNS matrix element cannot determined in this work and will be discussed in a separate work.

\acknowledgments
F. Xu was supported partially by NSFC  under Grant No. 11605076, as well as the Fundamental Research Funds for the Central Universities in China under the Grant No. 21616309.




\end{document}